\documentclass[twocolumn]{biophys}

\jno{kxl014}
\gridframe{N}
\cropmark{Y}
\doi{doi: 10.1529/biophysj.}

\def\o{\omega_{\mathrm{off}}}
\def\on{\omega_{\mathrm{on}}}
\def\F{\mathcal{F}}
\def\E{\mathcal{E}}
\def\kpas{K_{\mathrm{pas}}}
\def\pp{p_{\text{on}}/p_{\mathrm{off}}}
\def\A{\mathcal{A}}

\usepackage{amssymb,amsfonts,amsmath}
\usepackage{graphicx}
\usepackage{multirow}
\usepackage{longtable}
\usepackage {color}
\usepackage{setspace}
\usepackage{epsfig, epstopdf,latexsym,graphicx,bm,pifont}

\providecommand\bcdot{\boldsymbol{\cdot}}

\begin{document}

\title{Persistence of activity in noisy motor-filament assemblies}

\author{Raghunath Chelakkot,$^\dagger$ Arvind Gopinath,$^\ddagger$ L. Mahadevan $^{\dagger\ast}$}

\address{ $^\dagger$School of Engineering and Applied Sciences, Harvard University, Cambridge, MA 02138, USA; $^\ddagger$Max Planck Institute for Dynamics and Self-Organization, Goettingen 37077, Germany;  $^\ast$Department of Physics, Harvard University, Cambridge, MA 02138, USA.}

\begin{abstract}
{Long, elastic filaments cross-linked and deformed by active molecular motors occur in various natural settings. The overall macroscopic mechanical response of such a composite network depends on the coupling between the active and the passive properties of the underlying constituents and nonlocal interactions between different parts of the composite.  
In a simple one dimensional system, using a mean field model, it has been shown that the combination of motor activity and finite filament extensibility yields a persistence length scale over which strain decays.  {Here we study a similar system,  in the complementary limit of strong noise and moderate extensibility, using Brownian multi-particle collision dynamics-based numerical simulations that includes the coupling between motor kinetics and local filament extensibility.  While  the numerical model shows deviations from the mean field predictions due to the presence of strong active noise caused by the variations in individual motor activity,  several qualitative features are still retained. Specifically, for fixed motor attachment and detachment rates, the decay is length is set by the ratio of the passive elasticity to the active shear resistance generated by attached motors.} Our study generalizes the notion of persistence in passive thermal systems to actively driven systems with testable predictions.
}
{Submitted XXXX, and accepted for publication XXXX}
{*Correspondence: lm@seas.harvard.edu.\\
Address reprint requests to :\\
Editor
}
\end{abstract}

\maketitle

%\keywords{Motors/Decay length/Active elasticity}

%\abbreviations{SAM, self-assembled monolayer; OTS, octadecyltrichlorosilane}
\section*{Introduction}
The active strain generated by molecular motors moving on elastic filaments is  the principal mechanism of force generation and motion in cell biology \cite{Howard, Witman, Howard_2, Everaers, 
Heussinger, Igor, Grill, Vilfan}. In natural and reconstituted versions of such active composites, the localized strains due to motors deform the filaments dynamically, due to the interplay between elasticity, geometric constraints, active motor forces and noise due to fluctuations in the ambient medium and in motor activity.  These deformations in turn regulate the activity of the motors themselves~\cite{Howard_1,Brokaw_1}. A structurally ordered example of such a system is the eukaryotic flagellum that is made of relatively stiff filaments (microtubules), motors (dynein) and 
passive elastic elements (nexins) which together oscillate with well defined wavelengths and frequencies \cite{Witman, Sui_Downing, Machin, Brokaw_1,Mukundan}. While a variety of models of varying 
degree of complexity \cite{Howard_2, Everaers, Heussinger, Igor, Grill, Vilfan} are consistent with observations of increasing wavelength with increasing flagella length, experimental evidences  indicate 
that this relationship is altered for very long lengths \cite{Brokaw_2, Brokaw_3, Camalet, Hilfinger} and that the wavelengths attained in nature are self-limiting \cite{Witman, Howard_2}  even as 
the flagella themselves range from tens of microns to nearly a centimeter. This strongly  suggests that over large lengths, mechanical information transmission degrades substantially. 

In a passive context \cite{Heussinger}, it has been recently demonstrated  that when bundles of filaments are forced to bend, the shearing forces between them are mediated by extensibility, leading to a characteristic scale over which mechanical information is transmitted. In an active system such as an array of soft filaments driven by molecular motors, a possible consequence of this finite decay length of strain is that the weak extensibility of the filaments can limit the range of mechanical signal transmission between molecular motors and restrict the length scale over which motor coordination can occur. This will significantly affect the motor activity in filaments of lengths much larger than the strain decay length and could bring in spatial inhomogeneities in both strain and activity. As an example, in Fig. 1(a), we show two distinct  motor 
patches labeled I and II  cross-linking a pair of thin filaments that may bend and shear. For inextensible filaments, mechanical activity by group (I) and shear (sliding) induced by this patch is transmitted by 
attached motors in (II)  over arbitrarily  large inter-aggregate distances. For extensible filaments, there is an interplay between shear / slide and filament elongation along its contour length and leads to a 
decrease in the ability of group (I) to mechanically link with group (II).  
\begin{figure}[t]
\begin{center}
\includegraphics [width=\columnwidth] {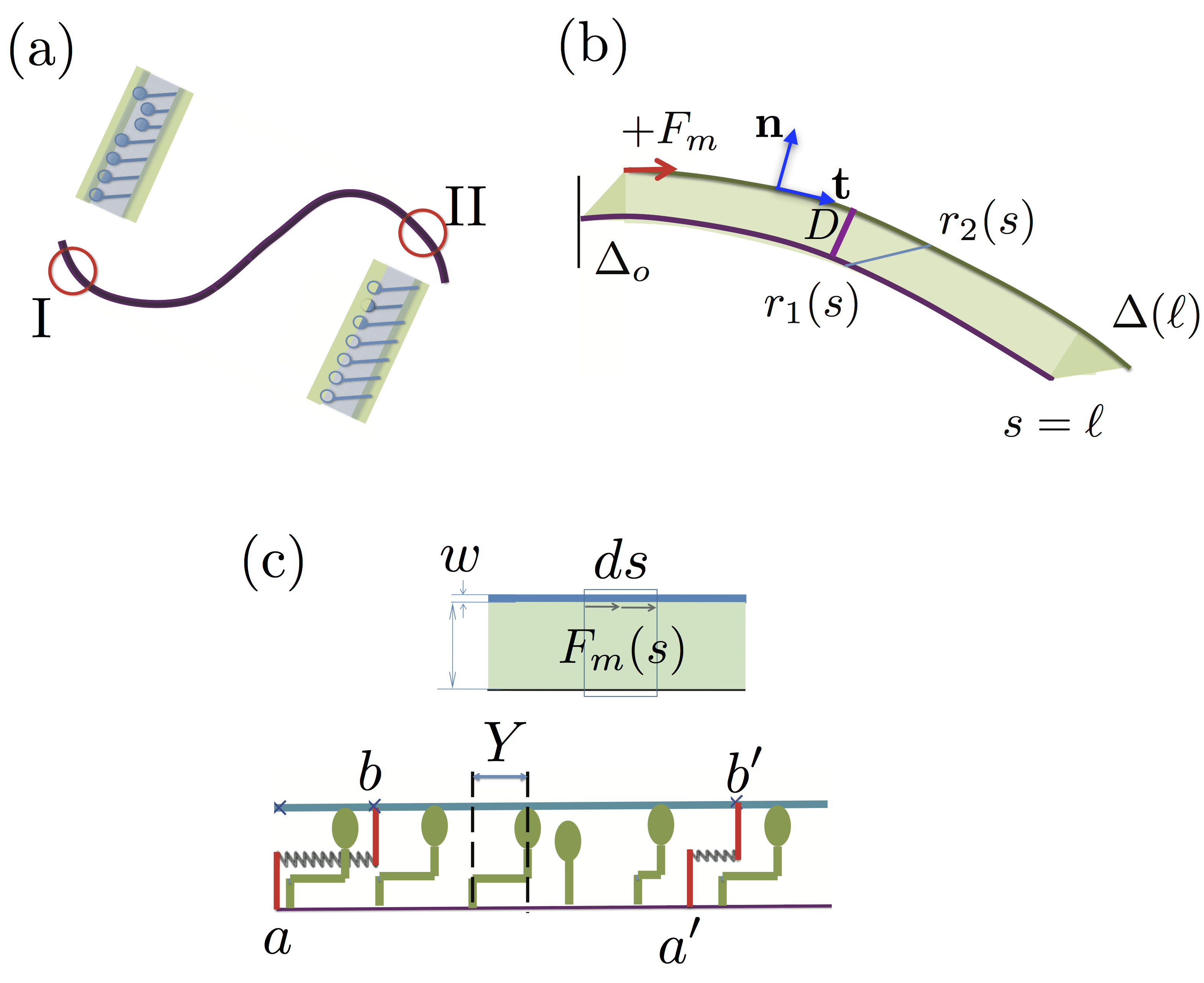}
\caption{\label{fig: 1}(a) {An active,  motor-filament composite illustrating two distinct motor aggregates,
groups of motors
(I) and (II).} Of interest is the distance between two motor groups beyond which they act as if isolated.  (b) Schematic and geometry of the passive filament composite with  length $\ell$, effective thickness $D+2w$. The arc-length is measured along the tangent ${\bf t}$. (c) {A sketch of the coarse-grained active gel (top) corresponding to the mean field approximation of
 the microscopic filament / motor system (bottom) illustrating the motors (blue) and cross-linking passive linkers (red).} }
\end{center}
\end{figure}

{ Using a mean field  approach valid in the noise-less limit ~\cite{Gopinath}, we recently  showed that the decay length of localized steady or oscillatory strains is determined by a combination of filament elasticity, passive shear resistance imposed by cross-linkers and the active viscoelastic properties of the motor aggregates. This analysis was performed by coarse-graining the stochastic nature of motor-filament interactions, and ignoring variations in motors kinetics due to the finite number of motors. However, even if thermal noise can be ignored  (such as when motors are rigidly fixed at the base and therefore do not diffuse) there is still an effective motor noise that arises due to discreteness of binding and unbinding events and fluctuations in the attachment times due to the finite number of motors. Moreover, there is a complete two way coupling between the motor activity and filament dynamics that arises as the motor activity causes filament strain, influences the collective motor dynamics. All these effects are not  captured in the mean field model.}

Here, we present a microscopic numerical model, in which we incorporate the effect noise arising from variations in motor kinetics, and implement the two-way coupling between motors and filament to study its influence on the persistence of active strain by comparing the results with the mean field predictions. { Predictions of the microscopic model  agree qualitatively with the mean-field results, despite the effect of strong active noise. However, motor noise is shown to result in richer dynamical features including modification of the motor duty ratio and localized regions of coherent oscillations due to the interplay between the local active shear resistance and filament elasticity.  }

\section*{Persistence of strain in a passive composite}
To understand the coupling between shear, bending and extension in a passive setting, we start with a minimal model of two elastic filaments of unstrained length $\ell$, lateral width $b$ and thickness $w \ll \ell $ held at a distance $D \ll \ell$ apart by a series of {passive, linearly} elastic springs with areal density {$\rho_{\mathrm{N}}$ and stiffness $k_{\mathrm{N}}$ (Fig.~\ref{fig: 1}(b)-(c)). The passive springs are compliant in shear { along the axial direction} but prevent any change in the distance between the filaments. 

At length scales large compared to the spacing between these springs, the composite acts as a filament of thickness $D$, with an effective shear modulus that depends on the spring stiffness, the density of springs, and the inter-spring spacing when  $b \gg {\mathrm{max}}(w,D)$.  By considering a sliding displacement of the top filament relative to the bottom due to a shear acting on its surface of area $b\ell$, with a stress $\sigma_{s}$, the net force is $\sigma_{s} b \ell$. Further, if the areal density of springs is $\rho_{\mathrm{N}}$, the shear is balanced by  $\rho_{\mathrm{N}} b \ell$ springs, all acting at the same time, and each contributing to shear resistance with spring constant $k_{\mathrm{N}}$. Assuming a shear (sliding) displacement of magnitude $\Delta$, force balance
\[
\Delta \: k_{\mathrm{N}} \: \rho_{\mathrm{N}} \: b \: \ell \sim \sigma_{s} \: b \ell \sim {G^{*}\Delta \over  D}\: b \: \ell 
\]
yields an effective shear modulus from these passive springs given by $G^{*} \sim D k_{\mathrm{N}}\rho_{\mathrm{N}}$. For a  strip of lateral width $w$, we have $G^{*} w/D  \sim w k_{\mathrm{N}}\rho_{\mathrm{N}}$ - consistent with previous analyses of a passive cross-linked railway track model \cite{Igor, Everaers}.

With this estimate in mind, we choose a local coordinate system characterized by an arc-length variable $s$ so that the filament is in the range  $s \in (0,\ell)$.  For two points ${\bf r}_{1}$ (on filament 1) and ${\bf r}_{2}$ (on filament 2) (Fig. \ref{fig: 1}(b)) which face each other when the filaments are in the undeformed state, we define the sliding deformation $\Delta (s)$ as difference between ${\bf r}_1$ and ${\bf r}_2$, relative to its initial difference before deformation. The angle made by the tangent to the centreline, $\theta(s)$ is then related  to the slide $\Delta(s)$ and the displacement $u(s)$ by geometry via the relations $d{\bf r}_{1}/ds  = {\bf t}$, $d{\bf t}/ds = {\mathcal{C}}{\bf n}$ and $ d{\bf n}/ds=-{\mathcal{C}}{\bf t}$, where ${\bf n}$ and ${\bf t}$ are the normal and tangent vectors, and ${\mathcal{C}}= d{\bf t}/ds \bcdot {\bf n} \approx \theta_{s}$ is the local curvature. In the deformed state, the position ${\bf r}_2$ relative to ${\bf r}_{1}$ can be written as, 
\begin{equation}
{\bf r}_{2} \approx {\bf r}_{1} + D\:{\bf n} + u \:{\bf t}.
\label{eq:SI2}
\end{equation}
Here we have assumed  $(D+w) \ll \ell$. The axial strain along the filament is $u_{s}$ where $d(.)/ds \equiv (.)_{s}$.
Using the relations
$
\Delta_{s} = D\left(|({\bf r}_{1})_{s}| -|({\bf r}_{2})_{s}|\right)$  and $
({\bf r}_{1})_{s} = {\bf t}$, we obtain, 
\begin{equation}
 ({\bf r}_{2})_{s} \approx {\bf t} - D\:{\mathcal{C}}\:{\bf t} + u_{s}\:{\bf t} + u \:\mathcal{C} \:{\bf n}. 
\label{eq:SI3}
\end{equation} 
For small deformations we can ignore quadratic/cross terms in the deformation to get
\begin{equation}
|({\bf r}_{1})_{s}| =1 , \:\:\:\:\:|({\bf r}_{2})_{s}| \approx 1 - D \theta_{s}  + u_{s} 
\label{eq:SI4}
\end{equation}
Substituting (\ref{eq:SI4}) into (\ref{eq:SI3}) we get
\begin{equation}
\Delta_{s} = D \theta_{s} - u_{s}. 
\label{eq:SI5}
\end{equation} 
Integrating (\ref{eq:SI5}) from $s=0$ to $s$ using the boundary condition $\theta(0)=0$ yields $\Delta (s) = D \theta(s) - u (s) $. 

The shape of the filament is then obtained by minimizing the energy $E_T$ of the composite due to bending, stretching and shear, with
\begin{equation}
E_{T} = b \int_{0}^{\ell} \left[\left( {{{{B_{\mathrm{pas}} \theta_{s}^{2} + K_{\mathrm{pas}} u_{s}^{2}}+ G_{\mathrm{pas}} \Delta^{2}}} \over 2}\right)
\right] \: d	s.
\label{eq:2}
\end{equation}
Here $B_{\mathrm{pas}} \sim E w^{3}$, $K_{\mathrm{pas}} \sim Ew $  and $G_{\mathrm{pas}}  \sim \rho_{\mathrm{N}}k_{\mathrm{N}}$   are the passive bending, stretching and shear moduli, with $E$ is the Young's modulus of the material. 
The Euler-Lagrange equations obtained by minimizing the functional (\ref{eq:2}) are given by ${\delta E_{T} / \delta {\Delta}} =0$ and ${\delta E_{T}/ \delta \theta} =0$ and lead to equilibrium static solutions. 

We consider solutions to (\ref{eq:SI5}) in two complementary limits. When the bending stiffness  $ B_{\mathrm{pas}} \rightarrow \infty$, the composite deforms due to shear and extension alone and $\Delta = -u$, so that the minimization of (\ref{eq:2}) provides
\begin{equation}
\Delta_{ss} - \left(G_{\mathrm{pas}} / K_{\mathrm{pas}} \right) \Delta = 0,
\label{eq-shear}
\end{equation}
which  yields the relaxation length scale for pure extension
$\ell^{2}_{E} \equiv {K_{\mathrm{pas}}/G_{\mathrm{pas}}} \sim {Ew / \rho_{\mathrm{N}}k_{\mathrm{N}}}.$
For inextensible filaments, $K_{\mathrm{pas}} \rightarrow \infty$, and $\Delta = {\it D}\theta$ and the minimization procedure provides the relation
$
\theta_{ss} - \left({\it D}^{2}G_{\mathrm{pas}} / B_{\mathrm{pas}} \right)\theta = 0,
%\label{eq-extension}
$
 thus yielding the persistence scale for pure bending $\ell^{2}_{B} \equiv  {B_{\mathrm{pas}} / {{\it D}^{2}G_{\mathrm{pas}} }} \sim  {Ew^{3} / {{\it D}^{2} k_{\mathrm{N}}\rho_{\mathrm{N}} }}$.
 
Allowing for sliding deformations at $s=0$ and keeping bending, extension as well as sliding terms, we repeat the minimization procedure to obtain the two coupled equations,  
\begin{eqnarray}
B_{\mathrm{pas}}\theta_{ss} + {\it D} K_{\mathrm{pas}}({\it D} \theta_{ss} - \Delta_{ss}) &=&0 \\
K_{\mathrm{pas}}(\Delta_{ss} - {\it D} \theta_{ss})-G_{\mathrm{pas}} \Delta &=& 0
\label{eq-bending}
\end{eqnarray}
In this general case, the second equation can rewritten as $ \Delta_{ss} -\ell_{*}^{-2}\Delta = 0$ and {indicates that variations in the sliding displacement}
$\Delta$ are associated with a persistence (decay) length
\begin{equation}
\ell_{*} \equiv  {{\ell_{B} \ell_{E}} \over {\sqrt{\ell^{2}_{B} + \ell^{2}_{E}}}} \sim {{w / D}\ \over \sqrt{1 + w^2/D^2}} \left({K_{\mathrm{pas}} /G_{\mathrm{pas}}}\right)^{1 \over 2}.
\label{eq:5}
\end{equation}
{Similar length scales appear in a variety of soft systems where bending, shear and extension are coupled, both in microscopic and macroscopic contexts~\cite{Reissner,Hatch}. }

To put these length scales in perspective, we look at typical parameters for a  flagellar axoneme. We take $w $ and $D$ as the radius and the spacing of microtubules,  $E$ as the Young's modulus of 
microtubules, $k_{\mathrm{N}}$  as the stiffness of passive (nexin) links and $\rho_{\mathrm{N}}$ as the density of these links.  Using $w \sim 20\:\text{nm}$, $D \sim 40\:\text{nm}$ \cite{Howard_1}, $E 
\sim 1.2$ GPa \cite{Howard_1, Howard_2}, $k_{\mathrm{N}} \sim 16-100$ pN $\mu$m$^{-1}$ \cite{Lindemann} and $w \rho_{\mathrm{\mathrm{N}}} \sim 10^{5}-10^7$ m$^{-1}$ \cite{Camalet,Lindemann}, 
we estimate $\ell_{E} \sim 200-500$ $\mu$m while $\ell_{*} \sim 80-200$ $\mu$m. This estimate is however with just the passive cross linkers contributing to the shear stiffness and therefore values should be treated as an upper limit. Motor activity will naturally influence this scale as attached motors contribute to the instantaneous shear resistance. Assuming that all motors are bound, with linear density $O(10^{8})$ m$^{-1}$  and effective spring stiffness of $10^{-3}$ N/m, we obtain $\ell_{E} \sim 5-10$ $\mu$m. Since sperm flagella are often much longer than this scale~\cite{Howard_2}, the role played by activity in modulating persistence lengths can be important. 

\section*{Mean-field model for an active composite}
For the case we treat here, corresponding to the case without bending wherein  the slide $\Delta(s) = -u(s)$ according to (\ref{eq:SI5}), motor activity modifies the decay length $\ell_{E}$ in the noiseless limit by changing the effective shear resistance in a manner controlled by motor kinetics. We being by summarizing the results of the mean field analysis~\cite{Gopinath}, valid in the noise-less limit with the focus on  systems with purely extensional modes of deformation. While the original analysis~\cite{Gopinath} allows for a more general form for the motor-filament interactions, here we consider a special case of the continuum theory consistent with  our simulations as shown in Fig.\ref{fig: 1}(b). 

Here, two cross-linked filaments are held together by a combination of active elastic motors and by passive elastic spring-like linkers. The active motors act as linear elastic springs of stiffness $k_m$,  and can either be attached to or detached from the filament. When the inter-link spacing  and the inter-motor spacing is much smaller than the filament length, the motor-filament assembly can be represented by a continuum description. In this case, for an inertialess filament, the sum of forces on the filament arising from filament extensibility, springs and motors must vanish. If $F_m$ is the active force due to a single attached motor and $\rho_m$ the motor density, this yields
\begin{equation}
(K_{\mathrm{pas}}u_{s})_{s} - G_{\mathrm{pas}}u + \rho_{m}F_{m} = 0,
\label{eq:6}
\end{equation}
which differs from previous system described by (\ref{eq-shear}) because of the active force density term $\rho_m F_m$.

{To estimate the active force density $F_m$ that contributes to an active shear resistance, we consider a minimal model for the motor-filament interaction. In this, active motors attach to the filament with zero extension (no pre-strain) and once attached walk relative to the filament with a velocity $v_m$. The motor velocity $v_m$ is linearly related to strain through a linear relation, $v_m=y_t=u_t+v_0(1-k_my/F_s)$, where $u_t$ is the velocity of the filament relative to the motor, $F_s$ is the motor stall force and $v_0$ is the zero load velocity of the motor.  Since one end of the motor is permanently attached to a rigid substrate, motor movement results in a motor extension $y$ and thereby causes a resisting force $k_m y$. }

Motor kinetics may be described by a set of population balances relating the attached and detached probability densities to the corresponding fluxes. Ensemble averaging the terms in these relations provide dynamical equations for the fraction of attached motors, $N(s,t)$ and the mean motor extension, $Y(s,t)$  (see ~\cite{Gopinath} for detailed derivation). The evolution equation  for $N$ is then given by
 \begin{equation}
N_{t} = {\omega}^{o}_{\mathrm{on}} (1-N)- {\omega}_{\mathrm{off}}N,
\label{eq-nt}
 \end{equation}
where $\on^o$ is the mean attachment rate of motors, and $\o$ is the load dependent detachment rate, defined as $\o=\o^{o} \F(\E,Y)$, where $\E \equiv k_{m}\delta^{2}_{m}/k_{B}T$ with $\delta_{m}$  being the extension corresponding to motor detachment, and the scaled mean extension of the attached motors, $Y \equiv \langle y \rangle /\delta_{m}$. Since the motor detachment is strain dependent, the evolution of $N$ is non-linearly coupled to the mean motor stretch $Y$,  as implied in (\ref{eq-nt}). Population balance relations then lead to an analogous evolution equation for $Y$  that is given by
\begin{equation}
{\small
{Y}_{t}  =  {u_{t} \over \delta_{m}} + {\mathcal{A}}_{1} {{\omega}^{o}_{\mathrm{off}}} \left({ {\mathcal{A}}_{2} }-Y\right)
+ {{\omega}^{o}_{\mathrm{on}}} Y \left({{1-N}
\over N}\right) 
}.
\label{eq-yt}
\end{equation}
In (\ref{eq-yt}), the first term on the right hand side is the stretch due to the passive motion of the attached motor, the second term gives the motor velocity relative to the filament, and the third term corresponds to the rate at which the mean strain changes due to the kinetics of motor attachment and arises from the difference in extension of attaching and detaching motors. {The two dimensionless parameters in (\ref{eq-yt})  $\A_1\equiv v_0k_m/(\o^o F_s)$ and $\A_2 \equiv F_s/(k_m\delta_m)$ relate the microscopic motor kinetics to the population averaged dynamics.} Once we have $N$ and $Y$, the motor force $F_m$ in (\ref{eq:6}) is written as
\begin{equation}
F_m = k_{m}\delta_{m}NY.
\label{eq:fm}
\end{equation}
Equations (\ref{eq:6})-(\ref{eq:fm}) provides a closed system for the the dynamics of filament displacement $u$, attached motor density $N$ and motor extension $Y$ in terms of dimensionless parameters $\A_1$, $\A_2$ and $\Psi=\o/\on$. These equations have a homogeneous stationary solution given by $N_{0}= (1+ \Psi {\mathcal F}_{0})^{-1}$ and $Y_{0} = {{\mathcal{A}}_{1} {\mathcal{A}}_{2}}({ {\mathcal{A}}_{1} + {\mathcal{F}}_{0}})^{-1}$, and a natural question is that of their stability to variations in the parameters.

Before we describe the global distribution of the strain field and the persistence of activity, we first examine the dynamics of a small fragment of the filament- motor composite of length $\ell_{s} \ll (K_{\mathrm{pas}}/G_{\mathrm{pas}})^{1 \over 2}$. In this limit, we can ignore filament extensibility and the motion of this segment relative to its neighbors can be mapped to that of a homogeneous population of motors acting on a rigid segment and working against an external  spring with effective stiffness $K_{\mathrm{s}}\propto G_{\mathrm{pas}}$.\footnote{Strictly speaking, there is also an active contribution to this effective spring constant that comes from attached motors in neighboring filaments -  but this just rescales $K_{s}$ and the mapping still holds. If $
\rho_{\mathrm{N}}=0$ then the passive part of $K_s \propto K_{\mathrm{pas}}$. } For such a fragment, when $K_{s} >0$,  a linear stability analysis of equations (\ref{eq-nt}) -(\ref{eq:fm})  reveals that the stationary states $N_0$ and $Y_0$ become linearly unstable and a stable oscillatory state emanate via supercritical Hopf-Poincare bifurcations\cite{Gopinath}.{ In the stable oscillatory state, the power input into the system due to activity balances the energy dissipated by motor viscosity. The effective motor friction characterizing this process is $\eta_{\mathrm{act}} \sim \rho_{m} N_{0}k_{m}(d\langle y \rangle /du_{t})_{u_{t}=0}$ where $\rho_{m}N_0$ is  the total  number of attached motors.} As $\Psi$ increases, the term $(d\langle y \rangle /du_{t})_{u_{t}=0}$ typically decreases, as does $N_{0}$ thus predicting an  increase in the frequency of the emergent oscillations $\omega_{c}$. 

{ To connect these localized solutions to the global extensional field in order to determine decay lengths of imposed extensions, we 
consider an active composite filament held fixed at one end ($s = 0$) while free at the other ($s=\ell$) which is forced harmonically.}  We assume that $u(0,t) = 0$ and $u(\ell,t)=u_{0}(\ell) + \epsilon \hat{u} {\mathrm{Real}}[e^{i\omega t}]$, where $\epsilon \hat {u}/u_0(\ell) \ll 1$ characterizes the deviation from the stationary extensional state  $u_0(s)$ which satisfies the equation
\begin{equation}
K_{\mathrm{pas}}  (u_{0})_{ss} - G_{\mathrm{pas}} u_{0} + G_{\mathrm{act}}C = 0,
\label{eq:steady}
\end{equation}
where $G_{\mathrm{act}}=\rho_mk_m$, the active analogue of passive shear modulus and $C \equiv \delta_{m} N_{0} Y_{0} = {\delta_m \over 1+\Psi {\mathcal{F}}_{0}} {{{\mathcal{A}}_{1} {\mathcal{A}}_{2}}  \over {{\mathcal{A}}_{1} + {\mathcal{F}}_{0}}}$. With boundary conditions $u_{0}(0)=0$ and $(u_{0})_{s}(\ell)=0$,  at leading order, i.e. $O(\epsilon^0)$  (\ref{eq:steady}) provides,
\begin{equation}
u_{0}(s)  =  \beta
  \left( 1 - \alpha \: e^{s \over \ell_{E}}
-   \left(1 - \alpha \right) \: e^{-{s \over \ell_{E}}}\right).
\label{eq:12}
\end{equation}
 Here, we note that the decay length $\ell_{E} =  \sqrt{ K_{\mathrm{pas}} / G_{\mathrm{pas}}}$, $\alpha = e^{-2{\ell \over \ell_{E}}}(1 + e^{-2{\ell \over \ell_{E}}})^{-1}$ and $\beta \equiv {{G_{\mathrm{act}} {\mathcal{A}}_{1} {\mathcal{A}}_{2} \delta_{m}} \over {G_{\mathrm{pas}} ({{\mathcal{A}}_{1} + \mathcal{F}_{0}}) (1+\Psi \mathcal{F}_{0})}}$ does not depend on the active shear resistance because motors sense only strain rates and not the actual strain. 

}
{At the next order, i.e. $O(\epsilon)$ we find the relations that determine the linearized response of  the system to externally imposed small amplitude perturbations. The magnitude of the time dependent part of the extension, $\hat{u}(s)$  due to the active force from the motors $\rho_{m} F_{m}$ can be found by substituting $\rho_{m}F_{m} = - G_{\mathrm{act}} \chi \hat{u}$ in (\ref{eq:6}) and yields
\begin{equation}
K_{\mathrm{pas}}\:\hat{u}_{ss} -  (G_{\mathrm{pas}} +  G_{\mathrm{act}}\chi)\:\hat{u} = 0.
\label{eq:osc}
\end{equation}
{where  $\chi(\omega)= - \delta_{m} (Y_{0}T_{2}+N_{0}T_{1})$  characterizes the complex linearized response of the system with 
$T_{1} \equiv - \left({i \omega \over \delta_{m}}
 \right)\left( i\omega + \omega^{o}_{\mathrm{on}}\Psi {\mathcal{F}}_{0}+ 
 \omega^{o}_{\mathrm{off}}{\mathcal{A}}_{1} \right)
 - \left({i \omega \over \delta_{m}}
 \right)
\left( \omega^{o}_{\mathrm{on}} \left({{Y_{0}} \over N_{0}^{2}}\right)
\left({{i \omega  + \omega^{o}_{\mathrm{on}}+ \omega^{o}_{\mathrm{off}}{\mathcal{F}}_{0}}
\over {\omega^{o}_{\mathrm{off}}{\mathcal{F}}'_{0}N_{0}}} \right)\right)^{-1} $
and  $T_{2} \equiv - \left({{i \omega  + \omega^{o}_{\mathrm{on}}+ \omega^{o}_{\mathrm{off}}{\mathcal{F}}_{0}}
\over {\omega^{o}_{\mathrm{off}}{\mathcal{F}}'_{0}N_{0}}} \right) T_{1}$\cite{Gopinath}.}
Seeking solutions to this equation of the form $\hat{u} \sim e^{s/\lambda}$ yields the following expression for the effective persistence length, 
\begin{equation}
\lambda_{E} \equiv  {\mathrm{Real}}[\lambda] = \sqrt{2}\: \left({{{\sqrt{\varphi_{1}^2 + \varphi_{2}^2} + \varphi_{1}}}}\right)^{-{1 \over 2}}
\label{lambda}
\end{equation}
where $\varphi_{1} = \left({G_{\mathrm{pas}}/ K_{\mathrm {pas}}}\right) +
\left({G_{\mathrm {act}} / K_{\mathrm {pas}}}\right) {\mathcal{R}}$ and $\varphi_{2} = \left({ G_{\mathrm act} / K_{\mathrm {pas}}} \right)\mathcal{I}$ { with ${\mathcal{R}}$ and ${\mathcal{I}}$ being the real and imaginary part of $\chi$ respectively}.  We note that in the absence of activity,  i.e. $G_{\mathrm act}=0$, we recover the expression $\lambda =\ell_E$. 

{In summary, the mean-field analysis predicts that  the relative importance of the active (motor) to passive (linker) elasticity is  controlled by $G_{\mathrm{act}}/G_{\mathrm{pas}}$, whereas in the absence of passive linkers the persistence length is controlled solely by $G_{\mathrm{act}}/K_{\mathrm{pas}}$. Further, (\ref{lambda}) implies that the persistence length is influenced by the imposed frequency $\omega$, though the frequency dependent complex response term, $\chi$  that controls the competition between intrinsic motor duty ratios and the extrinsic imposed frequency $\omega$ in determining the filament extension.}

\section*{Numerical model for the motor-filament composite}

\begin{figure*}[t]
\begin{center}
\includegraphics [width=2.1\columnwidth]{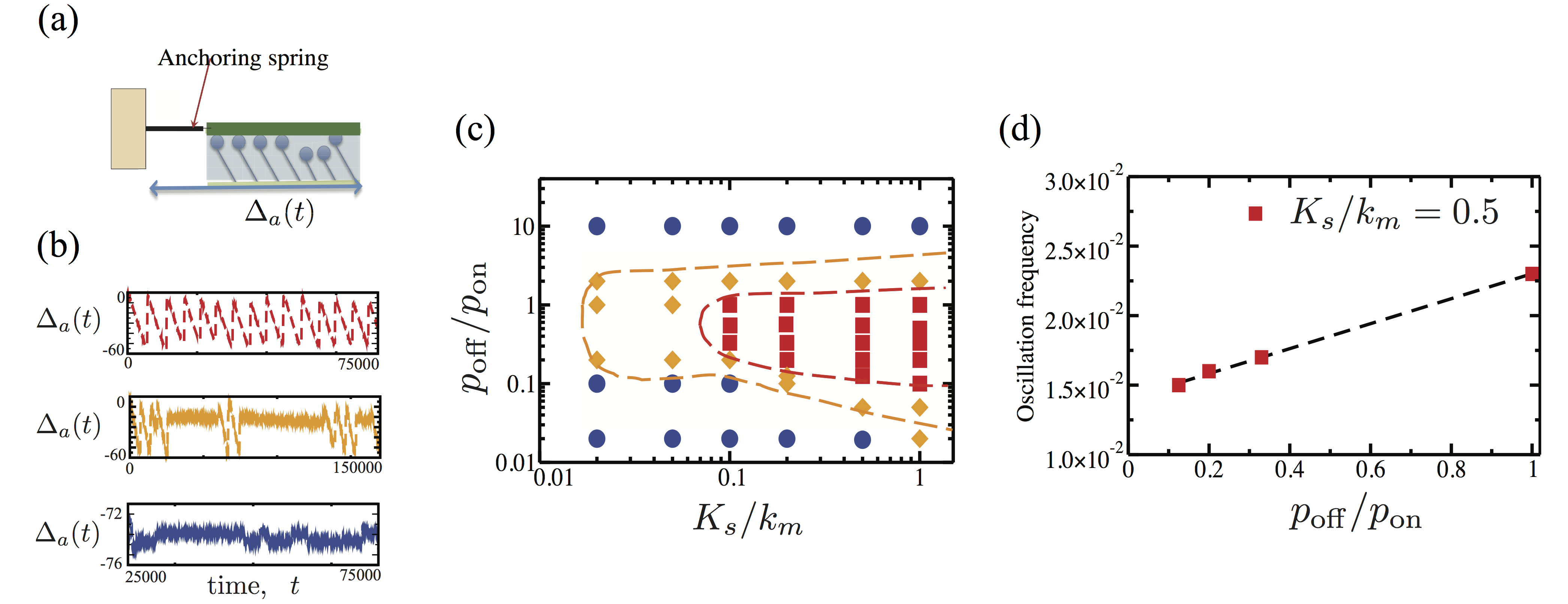}
\caption{\label{fig: 2}
(a) A schematic diagram of the filament-motor system for a small fragment.  The filament moves due to active forces imposed by attached motors - this motion is resisted by a anchoring spring of stiffness $K_{s}$.
(b) The time dependent displacement, $\Delta_a(t)/d_{m}$ for a fixed value for the ratio of connecting spring to motor stiffnesses, $K_{s}/k_m = 0.2$, and for zero load motor detachment/attachment probabilities, $p_{\mathrm{off}}/p_{\mathrm{on}} = 1.0$ (top),  $2$ (middle) and $10$ (bottom), illustrating various dynamical regimes observed in simulations.
(c) Phase diagram indicating the dynamical regimes as a function of the ratio $K_{s}/k_m$, and the ratio of zero load detachment-attachment probabilities ($p_{\mathrm{off}}/p_{\mathrm{on}}$).The observed dynamical regimes are, steady extension ($\circ$), intermittent oscillations ($\blacklozenge$), and steady oscillations ($\blacksquare$). Dashed curves are a guide to the eye.
 (d) The oscillation frequency, $\omega_c$ of a periodically oscillating filament with $K_{s}/k_m=0.5$, as a function of $p_{\mathrm{off}}/p_{\mathrm{on}}$.}
\end{center}
\end{figure*}

However, the mean field theory does not take into account the effect of local motor noise due to discrete attachment and detachment events and the finite number of motors. Also, there is no intrinsic coupling between filament extensibility and the mean attachment time of motors - as evidenced by the fact that $N_{0}$ and $Y_{0}$ are independent of $u_{0}$.
We now extend this minimal theory and investigate the role of increasing noise and fluctuations on the decay of both steady and oscillatory localized extensions.  

In our model, the continuous filament is made up of $N$ discrete spherical monomers of diameter $\sigma$ located at $r_{i}$, ($i=1,...,N$),
connected by an elastic  potential
\begin{equation}
u_{\ell}=\frac{\kappa_\text{l}}{2} \sum_{i=1}^{N-1} \left( |{\bf r}_{i+1}-{\bf r}_{i}| -b \right)^2.
\end{equation}
Bending stiffness is implemented via a three-body bending potential
\begin{equation}
u_\text{b} =\frac{\kappa}{2}\sum_{i=2}^{N-1} \left(\bf{t}_{i+1} -\bf{t}_{i}  \right)^2
\end{equation}
where
$
\bf{t}_i =({\bf r}_i-{\bf r}_{i-1})/|{\bf r}_i-{\bf r}_{i-1}|.
$
{ Here the effective bending rigidity $\kappa$ penalizes angular changes from a local straight geometry. We fix the filament length to be a constant $\ell=80\:\sigma$, and also fix the spring constants  $\kappa_\text{l}=\kappa=2\times10^4 \left(k_BT/\sigma^2 \right)$ in order to get a large filament persistence length $\ell_p\simeq250\:\ell$.} The bending and extensional stiffness are large enough so that the segment may be treated as a rigid inextensible segment. For simplicity and to { focus on the role of activity and noise}, we did not incorporate permanent passive cross links in the model. {With this  simplification, shearing is resisted  because of  the temporary crosslinks (bridges)  formed by attached motors.  These bridges disappear as motors detach at sufficiently high loads.} 
 
The active motors were modeled as linear springs with stiffness $k_{m}$ and equilibrium length $\ell_{m}$. The motors attach with zero mean strain and walk along the filament with discrete step size $d_m$.  As the motor heads move, the accompanying extension of the motor length  leads to a force ${\bf F}_{m}=-\kappa_m(\ell-\ell_m) \: \hat{\ell}$ that acts on the attached filament. {Thus, motor activity results in  localized active strain on the  section of the filament to which they are attached}. Consistent with previous experimental studies the load dependent stepping rate,  i.e. the stepping velocity of the motor head - is chosen to have the simple form $v_{m}(1-|{\bf F}_{m}|/f_{\mathrm{max}})$.  However, unlike in mean-field models where the attachment and detachment processes are coarse-grained into rates, in the simulations we prescribe {microscopic probabilities:} the load independent attachment probability$p_{\mathrm{on}}$,  is the probability of a detached motor to attach to the filament, while the motor detachment probability $\tilde{p}_\text{off}$ is defined by the piecewise function
\[
\tilde{p}_{\text {off}}=
\begin{cases}
p_{\text {off}}, & \text{if }|{\bf F}_m| \leq f_{\text {cut}}\\
1, & \text{if } |{\bf F}_m| > f_{\text {cut}}.
\end{cases}
\]
{This load dependence of the detachment probability of individual motors allows for the crucial two-way coupling between filament elasticity and the motor response and also sets the  critical motor extension at which detachment occurs. We take $f_{\mathrm{cut}}=f_{\mathrm{max}}$ in our simulations, although in general they are not equal.} Furthermore, the constant load independent detachment probability $p_{\text{off}}$ in conjunction with $p_\text{on}$ yields a well defined equilibrium attachment of motors at zero load, such that, $p_{\mathrm{on}}(1-
N)+p_{\mathrm{off}}N=1$. However, we note that despite the added noise, { our numerical model  is an approximate one in that  excluded volume interactions between the motors are not taken into account; these interactions will be significant only at large values of $p_{on}$ when almost all motors are always attached, a situation that we do not consider here}.
 
The position and velocity of the monomer beads are updated using velocity-Verlet algorithm, in  which the force acting on the monomer is calculated at every time step.  The entire system is immersed in a viscous medium using the Brownian multiple particle collision dynamics (MPC) scheme  which neglects the effects of inertia~\cite{Gompper}. According to this scheme, each monomer independently performs a stochastic collision with a phantom fluid particle with a momentum taken from the Maxwell-Boltzmann distribution  with variance $\rho k_{b} T$, where $\rho$ is the fluid density.

\begin{figure*}[t]
\begin{center}
\includegraphics [width=1.8\columnwidth]{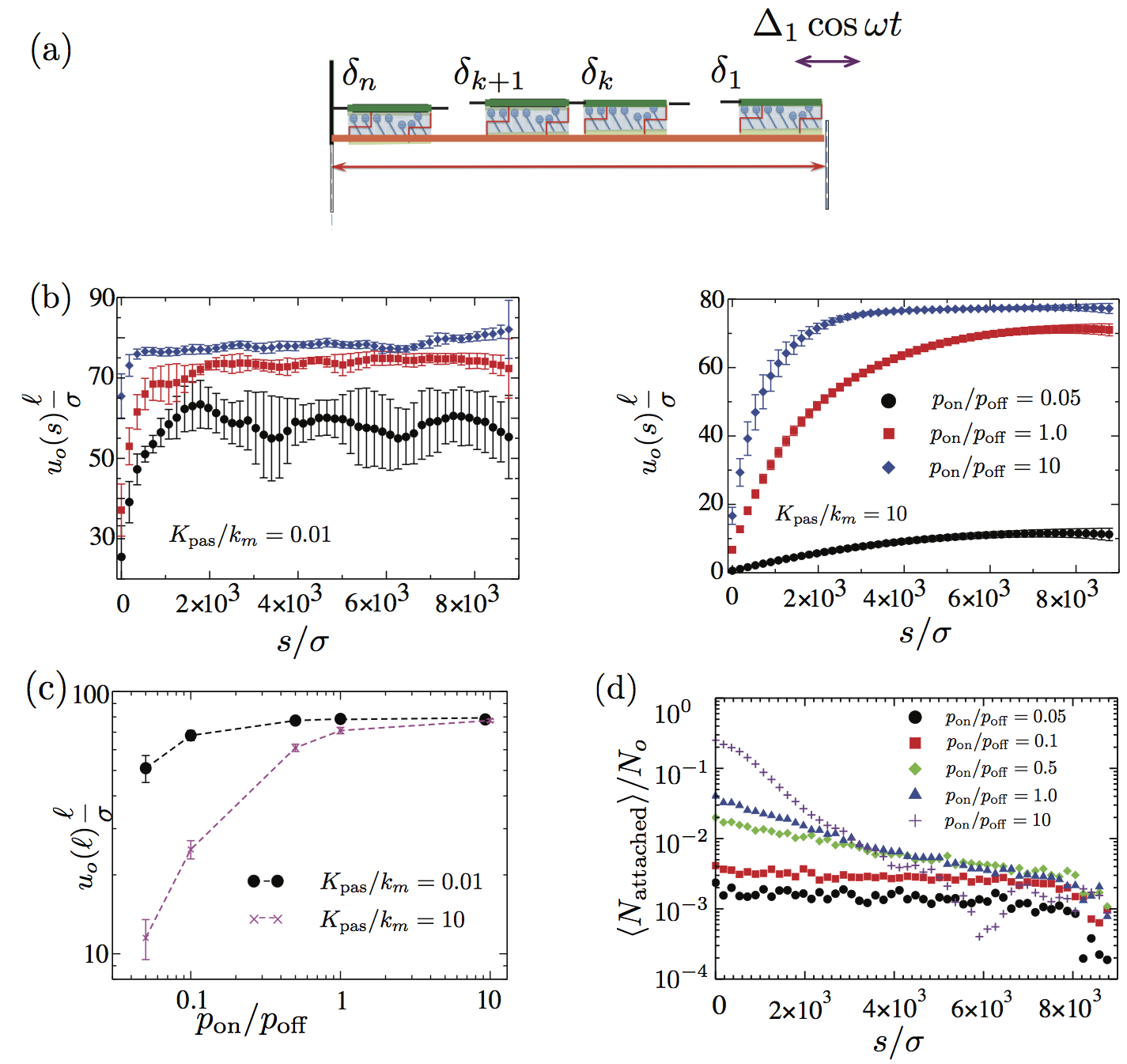}
\end{center}
\caption{\label{fig: 3} (a) Schematic sketch of a composite  filament made of 50 rigid segments, with $s$ as the coarse-grained position measured along the fiament.
(b)The mean steady state extension of material points as a function of their position from the clamped end ($s=0$). The vertical bars correspond to the root mean square deviation and  indicate the 
role of noise. The extension is less correlated for soft assemblies, when $K_{\mathrm{pas}}/k_{m} = 0.01$(left), while it is highly correlated when $K_{\mathrm{pas}}/k_{m} = 10$.
(c) Time averaged steady state extension  of the segment at the free end, $u_0(\ell)$, as a function of zero load motor attachment/detachment probability, $p_{\mathrm{on}}/p_{\mathrm{off}}$, for various 
values of passive spring stiffness, $K_{\mathrm{pas}}/k_m$. The extension increases with $p_{\mathrm{on}}/p_{\mathrm{off}}$, and decreases with $K_{\mathrm{pas}}/k_m$, consistent with  analytical 
predictions. For large   $p_{\mathrm{on}}/p_{\mathrm{off}}$ the extension saturates at a maximum value, set by the finite length of the filament-motor composite.
(d) The fraction of motors attached on a composite filament consists of 50 segments, as a function of their position from the clamped end for $K_{\mathrm{pas}}/k_m = 10$. The mean motor attachment is maximum close to the clamped end, where the local strain is relatively small.
 }
\label{fig4}
\end{figure*}

\begin{figure*}[t]
\begin{center}
\includegraphics [width=1.8\columnwidth]{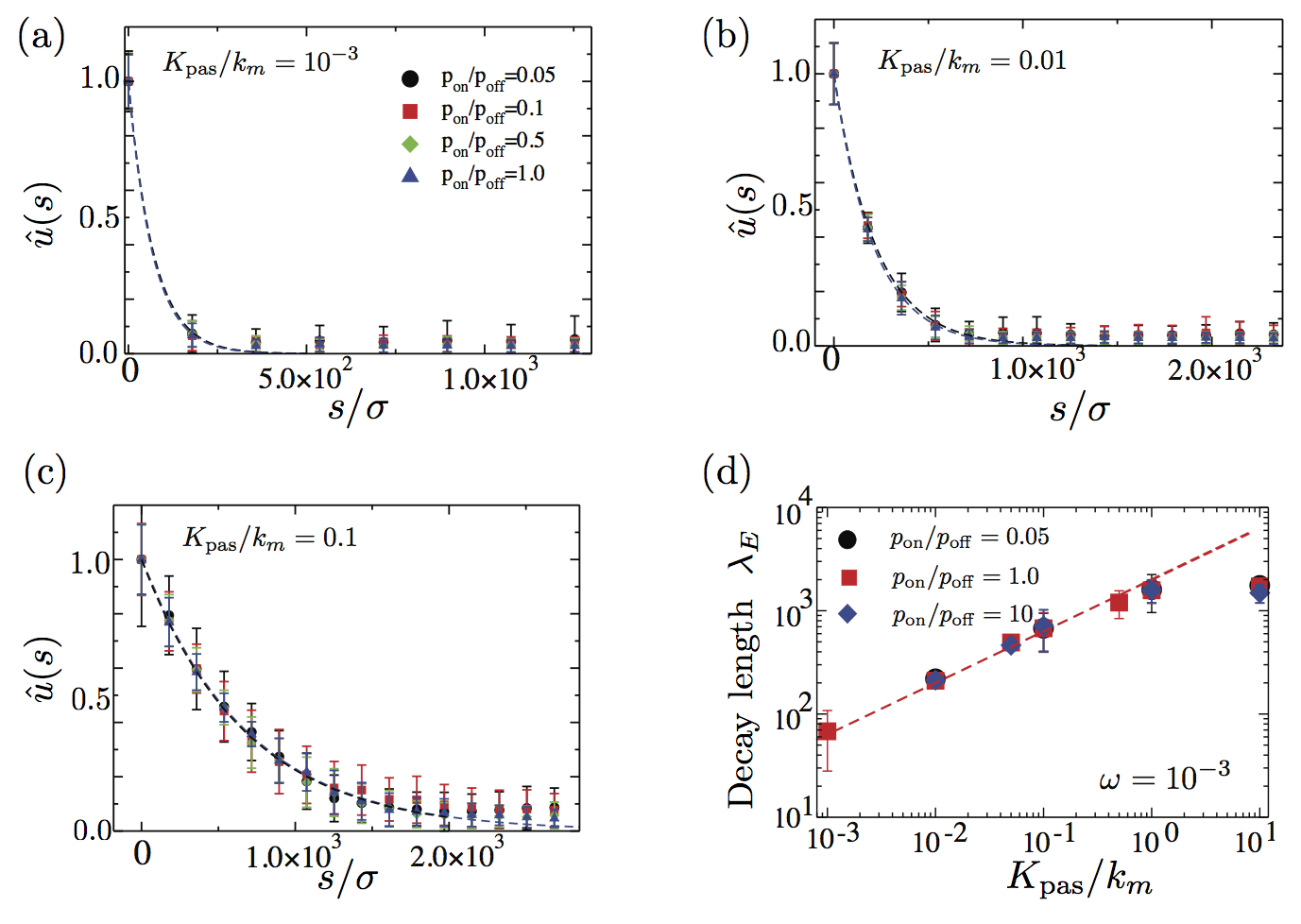}
\end{center}
\caption{\label{newfig}(a)-(c)The time averaged amplitude of imposed oscillations of the filaments as a function of  scaled positions, for various elasticity contrast (a) $\kpas/k_m = 10^{-3}$, (b) $10^{-2}$ and (c) $10^{-1}$. An exponential decay length is indicated whose value depends on both motor kinetics as well as the elasticity contrast. (d) The decay lengths obtained by analyzing  the amplitude of oscillations as a function of $K_{\mathrm{pas}}/k_m$ for various $p_{\mathrm{on}}/p_{\mathrm{off}}$.
}
\end{figure*}

{Using the numerical model,  we first analyze the effect of noise in dynamical regimes of a short inextensible fragment - this corresponds to the dynamics of filaments with length $\ell_{s}  \ll (K_{\mathrm{pas}}/G_{\mathrm{pas}})^{1 \over 2}$. This allows us to identify the way noise modified the predictions of the mean field model, which predicts a sharp separation between two states - a stable, stationary state and a stable, oscillatory state. We expect noise to make true boundary between these two regimes fuzzy.}

In Fig.~\ref{fig: 2}(a), we show a rigid filament anchored  to a wall via a linear elastic spring of stiffness $K_{s}$ that moves under the action of 800 motors.
%{Recall that the passive effective stiffness $K_{s} \propto G_{\mathrm {pas}}$ when $K_{\mathrm{pas}} = 0$ while  $K_{s} \propto K_{\mathrm{pas}}$ when $\rho_{\mathrm{N}} = 0$.}
Monitoring the {long time} displacement of the free end as a function of the spring constant $K_{s}$  and the probability ratio $p_{\mathrm{off}}/p_{\mathrm{on}}$ yields results summarized in Fig.~\ref{fig: 2}(b)-(d). The interplay between the softness of the filament composite $K_{s}/k_{m}$ and the activity $p_{\mathrm{off}}/p_{\mathrm{on}}$ yields three distinct dynamical regimes illustrated in Fig. \ref{fig: 2}(b). For fixed $K_{s}$ when $p_{\mathrm{off}}/p_{\mathrm{on}} \ll 1 $, there is a large force on the filament, and the free end of the filament attains 
a steady displacement with a well defined mean attached motor density, with fluctuations due to the noise imposed by discrete nature of motor binding and unbinding. For $p_{\mathrm{off}}/p_{\mathrm{on}} \gg 1$, the average number of motors attached to the filament is close to zero, and there is effectively no active force.  However there is an intermediate range in $p_{\mathrm{off}}/
p_{\mathrm{on}}$, where we observe first intermittent (Fig.\ref{fig: 2}(b)-middle) and then regular (Fig.\ref{fig: 2}(b)-bottom) filament oscillations. { While the regular oscillations are consistent with the mean field predictions, the intermittent oscillating state - where the filament-motor composite randomly switches from a stationary to an oscillatory state and back, is a new dynamical state observed only in the presence of noise.} A sweep of the parameter space yields the phase plot in Fig. \ref{fig: 2}(c). The parameter range over which steady oscillations  are seen is qualitatively consistent with theoretical predictions ~\cite{Gopinath}. However,  due to discrete motor noise, the boundary separating the steady and oscillatory states is no longer sharp - as expected, we observe a region of intermittent oscillations.  Within the oscillatory regime, the frequency increases with increasing $p_{\mathrm{off}}/
p_{\mathrm{on}}$ (Fig.~\ref{fig: 2}(d)), which also agrees with the trend seen from the mean field predictions of the theoretical model.

Next we study the role of {motor noise on} the extensibility of a composite filament that is made up of 50 equally sized rigid segments, each linked to adjacent neighbors by linear springs of stiffness $K_{s}$ (Fig. \ref{fig: 3}(a)) that account for the weak extensibility of the composite. Since no permanent cross links are present,  $\rho_{\mathrm{N}} = 0$, and  the effective passive stiffness  $K_{s} \propto K_{\mathrm{pas}} $.  We choose parameters for the filament elasticity and attachment and detachment probabilities such that the base state in the absence of imposed oscillations is steady (non-oscillatory) and stable and oscillate the free end at low frequency $(\omega \ll d_m/v_m)$, and small amplitude ($\Delta_{1} \ll \Delta_a$).  For sufficiently low frequencies, the results are qualitatively {independent of $\omega$} and hence we will focus on results obtained for a frequency $\omega = 10^{-3} (d_m/v_m)$. Quantifying the mean extension of the filament as a function of arc length $s$ of the extensible filament motor composite shown in Fig.~\ref{fig: 3}(b), we see that the maximum at the free end is influenced strongly by both $p_{\textrm{off}}/p_{\textrm{on}}$, and $K_{\mathrm{pas}}$. However, the discrete nature of motor activity causes large fluctuations in mean extension, especially for very soft filaments with $K_{\mathrm{pas}} /k_m \ll 1$ (Fig.\ref{fig: 3}b -left). Motor noise also results in uncorrelated spatial domains in extension for  $K_{\mathrm{pas}}/k_{m} = 0.01$ while a greater degree of correlation is observed for $K_{\mathrm{pas}}/k_{m} = 10$ (Fig.\ref{fig: 3}(b)-right). Concomitantly,  variations in the mean extension (shown as vertical bars) are also more correlated for stiffer filaments than for softer filaments. The noise decreases with increase in $p_{\textrm{on}}$, as shown in Fig. \ref{fig: 3}(b). 

Fig. \ref{fig: 3}(c) shows that the  steady state extension of the free end $u_{0}(\ell)/\sigma$ increases with $p_{\mathrm{on}}/p_{\mathrm{off}}$.  To compare the numerical predictions for steady state extension $u_{0}(s)$ to our mean field theory~\cite{Gopinath}, we simplify (\ref{eq:12}) by setting $\rho_{\mathrm{N}} = 0$, yielding the solution
\begin{equation}
u_{0}(s) = \left({G_{\mathrm{act}} 
\over K_{\mathrm{pas}}}\right) 
{{ {\mathcal{A}}_{1} {\mathcal{A}}_{2}} \over {{\mathcal{A}}_{1} + \mathcal{F}_{0}}} \left({ \delta_{m} \over {1+\Psi \mathcal{F}_{0}}}\right)
s(2 \ell - s)
\label{extension}
\end{equation}
from which one deduces that there  is no decay length scale  in the absence of permanent passive linkers. { We see from the right panel in Fig. \ref{fig: 3}(b) that this is needed the case.  While the simple theory predicts a quadratic form for the extension consistent with Fig. \ref{fig: 3}(b), we see that increasing the attachment probability yields a sharper gradient close to the clamped end and a flatter profile near the free end.} Setting  $G_{\mathrm{act}}=\rho_m k_m$, we find from (\ref{extension})  that the extension is an increasing function of  $k_m/K_{\mathrm{pas}}$ and is maximum at the free end, which are both
consistent with { our numerical results} (Fig.~\ref{fig: 3}(c)). {The dependence on the activity $\Psi$ is also qualitatively captured as the slope increases with the ratio $p_{\mathrm{on}} /p_{\mathrm{off}}$. } The extension at the free end $u_0(\ell)$ obtained from (\ref{extension}) indicates that $u_0(\ell)$ decreases with $\kpas/k_m$ while it increases with $(\pp)$ - both these trends are observed in the numerical simulations(Fig.\ref{fig: 3}(d)).

Since the motor detachment is strain dependent, the motor imposed extensions in turn regulate the motor kinetics - {and specifically modify the mean attached time}. In Fig. \ref{fig: 3}(d) we plot the mean fraction of motors attached, $\langle N_{\text{attached}}\rangle/N_0$,  as a function of the distance from the anchored end, $s$ for $\kpas/k_m=10$. When $\pp \lesssim 0.1$ the motor imposed extension of the filament is weak due to less number of motors attached, and the distribution of attached motors along the composite filament is approximately uniform. However for large values of $\pp$ ($\pp \gtrsim 1$) the enhanced motor attachment imposes significant extension to the filament, which is maximum at the free end. The nonuniform extension of the composite leads to a nonuniform distribution of attached motors (Fig.\ref{fig: 3}(d)), where the mean motor attachment is maximum near the anchored end ($s=0$) and decreases with the distance from that point, as the filament strain increases. This leads to a diminished motor activity near the free end. {These features are not seen in the mean field predictions and are thus a direct consequence of the noise due to the discrete nature of motor binding and unbinding events.}

{We finally calculate the length-scale corresponding to the decay of the amplitude of the imposed oscillatory strain ($\hat{u}$)  along the composite, for a range of $\kpas/k_m$ and $\pp$. For all values of $\kpas/k_m$  and $\pp$, our analysis show that $\hat{u}$ decays exponentially with  distance $s$ from the free end, as shown in Fig.~\ref{newfig}(a)-(c). Further, the decay length of the exponential  increases when $\kpas/k_m$ is varied from $10^{-3}$(Fig.\ref{newfig}(a)) to $10^{-2}$(Fig.\ref{newfig}(b)) and $10^{-1}$(Fig.\ref{newfig}(c)). An exponential fit to the amplitude along the composite for different parameter values allows us to estimate the decay length $\lambda_E$ as a function of $s$ (Fig.\ref{newfig}(d)). This  indicates that for a fixed $\pp$,  $\lambda$ is scales as $(\kpas/k_m)^{1\over2}$.  However, the decay length exhibits only weak dependence on motor kinetics for the 
range of attachment probabilities investigated. We also find that the value saturates to a constant in simulations with large $K_{\mathrm{pas}}/k_m$, which we ascribe to  finite system size effects.
 
To compare with theoretically predicted decay length to our simulation results, we set 
$\rho_{\mathrm{N}} = 0$  in (\ref{lambda}) to obtain the decay length
\begin{equation}
\lambda_{E} =
\left(2{K_{\mathrm{pas}} \over k_m \rho_m}\right)^{ { 1 \over 2}}
\left[ { 1 \over  \sqrt{ |\chi| + {\mathrm{Real}} (\chi)}}\right]
\label{eq:18}
\end{equation}
Since $\chi$ is a function of imposed frequency and motor kinetics, and is to leading order independent of filament stiffness, the term in the square brackets is a constant for a given $\omega$ and $\pp$. As shown in Fig.\ref{newfig}(d), this prediction agrees very well with our simulations.  }
\section*{Conclusions}

The scale over which active deformations persist in a fluctuating environment are basic questions in the study of living matter.  Here, we address both these questions in the context of an ordered composite structure consisting of elastic elements that can be stretched and/or sheared  by motors that can bind to and unbind from them, taking to account the individual motor kinetics.
 Even though extensibility is negligible locally, it affects the scale over which deformations persist owing to the competition between extensibility and shear.  In particular, for active systems, we show that the emergence of an oscillatory instability in a short segment leads to deformations that do not persist forever and instead decay over longer length scales. Further, the finite number of motors attached causes an intermediate regime of intermittent oscillations, which was not predicted by models that ignore active noise.  Our analysis shows that the feedback due to the ability of attached motors to sense local extension rates influences the motor activity and modifies the effective length scale over which strain decays.  

Our simulations suggest that deviations from the mean field theory - even for weak extensibility -  can arise in the limit of very small number of attached motors  (or small attachment rates) and very large 
number of attached motors (noise acting cumulatively at large attachment rates).  We find that, for a given motor activity, the decay length of strain predicted by the mean field theory, $\lambda_E \sim 2 (K_{\mathrm{pas}}/
G_{\mathrm{pas}})^{1\over 2}$, is valid even in the presence of noise. This confirms a finite range of correlated activity that might be relevant for natural examples of ordered active matter such as eukaryotic flagella~
\cite{Woolley} by setting a limit over which mechanical coordination can persist. It also raises interesting questions about how to generalize our analysis to disordered mixtures of motors and cytoskeletal 
filaments ~\cite{Dogic-2011,Dogic-2012}, particularly in the limit when the coordination number is small and when it approaches the isostatic limit.  

%\begin{acknowledgments}
%We acknowledge funding for this research provided by the MacArthur Foundation (LM), and computational support from the Brandeis HPC. 
%\end{acknowledgments}
%\bibliographystyle{biophysj}
%\bibliography{persistence}{}

%\end{article}
\end{document}